\documentclass[11pt]{article} 
\usepackage{a4wide} 

\RequirePackage[latin1]{inputenc}     
 
\usepackage[english]{babel} 
\usepackage{amsmath} 
\usepackage{amsfonts} 
\usepackage{amssymb} 
\usepackage{xspace} 
\usepackage{latexsym} 
\usepackage{url} 
\usepackage{xspace} 
\usepackage{fancyvrb} 
\usepackage[all]{xy} 

\usepackage{graphics,color}              
\RequirePackage[latin1]{inputenc}   
 
\newenvironment{restate-proposition}[2][{}]{\noindent\textbf{Proposition~{#2}}\;\textbf{#1}\  
}{\vskip 1em} 
 
\newenvironment{restate-theorem}[2][{}]{\noindent\textbf{Theorem~{#2}}\;\textbf{#1}\  
}{\vskip 1em} 
 
\newenvironment{restate-corollary}[2][{}]{\noindent\textbf{Corollary~{#2}}\;\textbf{#1}\  
}{\vskip 1em}

\newcommand{\Proofitem}[1]{\medskip \noindent $#1\;$}

\newcommand{\hbra}{\noindent\hbox to \textwidth{\leaders\hrule height1.8mm depth-1.5mm\hfill}} 
\newcommand{\hket}{\noindent\hbox to \textwidth{\leaders\hrule height0.3mm\hfill}} 
\newcommand{\ratio}{.3}

 \newtheorem{theorem}{Theorem}

  \newtheorem{corollary}[theorem]{Corollary} 
  \newtheorem{proposition}[theorem]{Proposition}

 \newcommand{\qed}{\hfill${\Box}$}

\newcommand{\Figbar}{{\center \rule{\hsize}{0.3mm}}}    
 
\newcommand{\cl}[1]{{\cal #1}}          
\newcommand{\la}{\langle}               
\newcommand{\ra}{\rangle}

\newcommand{\ol}[1]{\overline{#1}}      
 
\newcommand{\Gives}{\vdash}             
\newcommand{\Givesa}{\vdash^{a}}        
\newcommand{\Givespi}{\vdash^{\pi}}     

\newcommand{\arrow}{\rightarrow}        
\newcommand{\tarrow}{\stackrel{+}{\rightarrow}}        
 
\newcommand{\lets}[3]{{\sf let} \ #1 = #2 \ {\sf in} \ #3}    
\newcommand{\letv}[3]{{\sf let} \ (#1 = #2)^* \ {\sf in} \ #3}    

\newcommand{\mlets}[4]{{\sf let}_{#1} \ #2 = #3 \ {\sf in} \ #4}    
\newcommand{\mletv}[4]{{\sf let}_{#1} \ (#2 = #3)^* \ {\sf in} \ #4}    

\newcommand{\pr}{\parallel}             
\newcommand{\Alt}{ \mid\!\!\mid  } 
\newcommand{\isum}{\oplus} 
 
\newcommand{\infer}[2]{\begin{array}{c} #1 \\ \hline #2 \end{array}} 
 
\newcommand{\inter}{\cap}               
 
\newcommand{\w}[1]{{\it #1}}    

\newcommand{\s}[1]{{\sf #1}}    

\newcommand{\set}[1]{\{#1\}}

\newcommand{\adm}[1]{\underline{#1}}   
\newcommand{\rb}[1]{#1^{o}}      
\newcommand{\cpst}[1]{\overline{#1}}   
\newcommand{\tname}[1]{{\cl N}(#1)}      
\newcommand{\cname}[1]{{\it Ch}(#1)}      

\begin{document}

\title{A decompilation of the $\pi$-calculus\\ and its application to termination}


\author{Roberto M. Amadio\\ 
Universit\'e Paris Diderot\\ {\small (UMR CNRS 7126)}}

 \maketitle 

\begin{abstract}
We study the correspondence between 
a concurrent lambda-calculus in administrative, continuation passing style
and a pi-calculus and we derive 
a termination result for the latter.

\end{abstract}

\section{Introduction}
There are two complementary explanations of the $\pi$-calculus.
The first one is to regard it as an extension of a rather standard
process calculus such as CCS while the second one is to present 
it as an intermediate language for compiling higher-order languages including
various imperative/concurrent extensions of $\lambda$-calculi and object-oriented calculi.
The first view was put forward in the original presentation \cite{MPW92}
and explains the transfer of effective operational semantics techniques
from CCS to the $\pi$-calculus. The second view  
gradually emerged through a series of encodings starting from, {\em e.g.},
\cite{M92} and it explains the expressivity of the calculus while providing 
guidance in selecting its essential aspects.

Taking the second view, it has been stressed (see, {\em e.g.},
\cite{Boudol97}) that the translations from the $\lambda$-calculus to
the $\pi$-calculus can be understood as the {\em composition} of two
familiar compilation techniques.  In the first step, the
$\lambda$-term is put in an {\em administrative form} (AF) where all
values are explicitly named and in the second one a {\em continuation passing style (CPS)
translation} is applied so that the evaluation contexts are passed
explicitly as an argument. 

We note that neither the notion of administrative
form nor that of CPS translation are canonical.
Our purpose here is to provide a concrete presentation of this 
approach for a call-by-value $\lambda$-calculus and then 
for a parallel and concurrent extension of it. 
We show that the two compilation steps 
commute nicely with the reduction relations and the typing
disciplines. Moreover, we identify languages
which contain the image of the compilation and are isomorphic to
natural fragments of the $\pi$-calculus.
The situation is summarized in table \ref{overview-table} where the
$\lambda$-calculi in administrative form play a prominent role since
on one hand the ordinary $\lambda$-calculi can be regarded as a {\em
retraction} of the administrative ones (symbol $\triangleleft$) and
on the other hand they contain sub-calculi (symbol $\supset$) in CPS
form which are isomorphic (symbol $\cong$) to $\pi$-calculi.
Also it is at this level, that it is natural to introduce a notion of 
concurrent access to a resource, {\em i.e.}, a definition.
In this framework, one can {\em decompile} terms of the $\pi$-calculus
into familiar $\lambda$-terms. As an application of the
correspondence, we show that the termination of a (concurrent)
$\lambda$-calculus entails the termination of a corresponding
(concurrent) $\pi$-calculus.
Section \ref{fun-sec} covers the functional case,
section \ref{conc-sec} generalizes it to  the parallel and concurrent case,
and section \ref{pi-sec} makes explicit the correspondence with the
$\pi$-calculus. Omitted specifications, some concurrent programming examples, 
and  proof sketches of the main results are available in appendices \ref{spec-app},
\ref{expressivity-sec},  and \ref{proofs-sec}, respectively.

\begin{table}
{\footnotesize
\[
\begin{array}{|c|ccccccc|}

\hline
&\lambda-\mbox{notation}   &        &\mbox{Adm. Form (AF)}   &      &\mbox{AF in CPS style}  &  &\pi-\mbox{notation} \\\hline   

\mbox{{\sf Functional}} &\lambda        &\triangleleft

&\lambda^a           &\supset    &\lambda^{ak} &\cong  &\pi^f\\

\inter & & & & & &  &                              \\

\mbox{{\sf Concurrent}}
&\lambda_{\pr}  &\triangleleft     &\lambda^{a}_{\pr}   &\supset    &\lambda^{ak}_{\pr}   &\cong &\pi \\\hline

\end{array}
\]}
\caption{Overview of calculi and their relationships}\label{overview-table}
\end{table}

\paragraph{Related work}
This work arises out of a long term effort of presenting the
$\pi$-calculus to an audience familiar with the $\lambda$-calculus but
not necessarily with process calculi.  We tried this first in the
context of a book \cite{AmadioCurien97} and then more recently in the
context of a graduate course \cite{Amadio10}.  The application to
termination appears as a natural test for the presented compilation
techniques and can be regarded as a natural continuation of recent
work on the termination of (higher-order) concurrent calculi.
In this respect, we find it remarkable that in the presented approach one can drop completely the
notion of {\em stratified region} \cite{Boudol07,Amadio09,T10,DHS10}. As far
as the $\pi$-calculus is concerned, we believe the presented approach complements
those described in \cite{YBH04,S06} in that we reduce the termination of 
a fragment of the $\pi$-calculus to the termination of the $\lambda$-calculus 
by `elementary' means. In particular, we do not need to develop reducibility candidates/logical 
relations techniques for the $\pi$-calculus, nor do we require specific knowledge
of the operational semantics of the $\pi$-calculus.

\paragraph{Requirements}
The reader is supposed to be acquainted with the simply typed $\lambda$-calculus, see, {\em e.g.},
\cite{Girard89}, its evaluation strategies and continuation passing style translations, 
see, {\em e.g.}, \cite{Plotkin75}, and to have some familiarity with the syntax of the
$\pi$-calculus, see, {\em e.g.}, \cite{Milner99} and its reduction semantics. We shall make no use
of the so called labelled transition systems and related notions of bisimulation.

\paragraph{Renaming}
In the following calculi, all terms are manipulated 
up to $\alpha$-renaming 
of bound names. Whenever a structural congruence
or a reduction rule is applied, it is assumed that terms have been
renamed so that all binders use distinct variables and these variables are distinct from
the free ones.  Similar conventions are applied when performing a substitution,
say $[T/x]T'$, of a term $T$ for a variable $x$ in a term $T$'.
We denote with $\w{FV}(T)$ the set of variables occurring free in a term $T$.

\paragraph{Syntax, Structural congruence, Reduction, and Typing}
For each calculus, we specify the syntactic categories, the notion of structural  congruence, the reduction rules, and the typing rules.
In all calculi, we assume a syntactic category $\w{id}$ of identifiers (or variables) which we denote with $x,y,z,\ldots$.
Structural congruence is the least equivalence relation, denoted with $\equiv$, which is 
induced by the  displayed equations,  $\alpha$-renaming (which is always left implicit), and
closed under the operators of the language.
The reduction relation is the least binary relation $\arrow$ that includes the pairs given
by the rewriting rules and such that $M \arrow N$  if $M\equiv M'\arrow N'\equiv N$.
The basic judgment of the typing rules assigns a type to a term in a context $\Gamma$.
The latter is a function mapping a finite set of variables to types. When writing $\Gamma,x:A$
it is assumed that $x$ is not in the domain of definition of $\Gamma$. In the parallel/concurrent
extensions we distinguish between value types and types. The latter are composed of value types 
plus a distinct behavior type $b$. Terms of a behavior type do not return a result. As such 
a behavior type cannot occur in a context or as the type of the argument of a function.

\paragraph{Vectorial notation}
We shall write $X^+$ ($X^*$) for a non-empty (possibly empty)
finite sequence $X_1,\ldots,X_n$ of symbols.
By extension, $\lambda x^+.M$ stands for $\lambda x_1\ldots\lambda x_n.M$,
$A^+\arrow B$ stands for $(A_1\arrow \cdots \arrow (A_n\arrow B) \cdots )$,
$[V^+/x^+]M$ stands for $[V_1/x_1](\cdots [V_n/x_n]M \cdots)$,
$x^+:A^+$ stands for $x_1:A_1,\ldots,x_n:A_n$, 
$\Gamma \Gives M^+:A^+$ stands for $\Gamma \Gives M_1:A_1,\ldots, \Gamma \Gives M_n:A_n$, and
$\s{let} \ (x=V)^+ \ \s{in} \ M$ stands for 
$\lets{x_{1}}{V_{1}}{\cdots \lets{x_{n}}{V_{n}}{M}}$.

\section{The functional case}\label{fun-sec}
In this section we explore the functional case, namely the upper part of table
\ref{overview-table}. This has at least two advantages: first one can get an idea of
the approach in a simple familiar framework and second it clarifies
the additions to be made to achieve parallelism and concurrency.

\begin{table}
{\footnotesize
\begin{center}
{\sc Syntax}
\end{center}
\[
\begin{array}{lll}

V &::= * \Alt (\lambda \w{id}.M) \Alt \w{id} &\mbox{(values)}\\
M &::= V \Alt (MM)                           &\mbox{(terms)} \\
E &::= [~] \Alt E[[~]M] \Alt E[V[~]]         \quad&\mbox{(eval. contexts)} \\
A &::= 1 \Alt (A\arrow A)                    &\mbox{(types)} 
\end{array}
\]
\begin{center}
{\sc Reduction rule}
\end{center}
\[
\begin{array}{c}
(\beta_{V}) \qquad E[(\lambda x.M)V] \arrow E[[V/x]M]
\end{array}
\]
\begin{center}
{\sc Typing}
\end{center}
\[
\begin{array}{cc}

(\w{id})\quad \infer{x:A \in \Gamma}
{\Gamma \Gives x:A} 

&(*)\quad \infer{}
{\Gamma \Gives *:1} \\ \\ 

(\lambda)\quad \infer{\Gamma,x:A \Gives M:B}
{\Gamma \Gives \lambda x.M:A\arrow B}

&(@)\quad \infer{\Gamma \Gives M:A\arrow B\quad \Gamma \Gives N:A}
{\Gamma \Gives MN:B}

\end{array}
\]}
\caption{A simply typed, call-by-value  $\lambda$-calculus: $\lambda$}\label{lambda-def}
\end{table}

\paragraph{A $\lambda$-calculus}
To start with we introduce in table
\ref{lambda-def} a standard, simply typed, call-by-value
$\lambda$-calculus ($\lambda$).  We specify, syntax, reduction and
typing rules. We follow this pattern in the following calculi too,
possibly adding the specification of a structural congruence.

\paragraph{A $\lambda$-calculus in administrative form}
A corresponding calculus in {\em administrative form} ($\lambda^a$) is presented in table \ref{lambda-anf}.
The basic idea is to attribute a name to each value and to compute by replacing names with
names. Notice that in $\lambda^a$ we restrict the values in a \s{let} definition
to be either abstractions or constants `$*$'. 
Beyond values and terms, we introduce a new syntactic category of `declarations'
which are terms possibly preceded by a list of value declarations.
Strictly speaking the application only applies to terms, however if 
$D_i =\letv{x_i}{V_i}{M_i}$, for $i=1,\ldots,n$, then we regard $@(D_0,D_1,\ldots,D_n)$ as
an abbreviation for $\letv{x_0}{V_0}{\cdots \letv{x_n}{V_n}{@(M_0,\ldots,M_n)}}$ 
(the order of the declarations is immaterial up to structural congruence).
Similar care is needed when substituting a declaration $D =\letv{x}{V}{M}$ in an evaluation context 
$E$. We remark that $E$ can be written as $\letv{x'}{V'}{E'}$ where $E'$ does not start
with a $\s{let}$ declaration. Then by $E[D]$ we mean the declaration
$\letv{x}{V}{\letv{x'}{V'}{E'[M]}}$ where it is intended that 
the names $x'^*$ do not occur free in $E'$. \footnote{There 
is an alternative presentation 
where declarations and applications can be inter-mixed freely; 
we found the current presentation more handy for our purposes.}

We use $\tname{A}$ for the type of names carrying values of type $A$; 
as the reader might have guessed these names correspond 
to the channel names of the $\pi$-calculus.
The $\lambda^a$-calculus is {\em polyadic} in the sense that the application may
have any finite, positive number of arguments. This choice allows to represent 
directly the following CPS translation as a translation from the $\lambda^a$ calculus
to a fragment of the $\lambda^a$ calculus in `CPS form' and moreover the latter corresponds
directly to a {\em polyadic} $\pi$-calculus.
Nevertheless, we notice that 
the $\lambda^a$-calculus contains a {\em monadic sub-calculus} that
we denote with $\lambda^{am}$ where the types
are restricted as follows:
\[
A::= \tname{1} \Alt \tname{A\arrow A} \qquad (\mbox{monadic types})
\]
This sub-calculus is closed under reduction (provided 
we consider typable terms) and
it suffices to encode the simply typed $\lambda$-calculus.

\begin{table}
{\footnotesize
\begin{center}
{\sc Syntax}
\end{center}
\[
\begin{array}{lll}

V &::= * \Alt (\lambda \w{id}^{+}.D)             &\mbox{(values)}\\
D &::= \letv{\w{id}}{V}{M}                       &\mbox{(declarations)} \\
M &::= \w{id} \Alt  @(M,M^{+})                   &\mbox{(terms)} \\
E &::= \letv{\w{id}}{V}{[~]} \Alt E[@(\w{id}^*,[~],M^*)] \quad &\mbox{(eval. contexts)} \\
A &::= \tname{1} \Alt \tname{A^+\arrow A}            &\mbox{(types)} 
\end{array}
\]
\begin{center}
{\sc Structural Congruence}
\end{center}
\[
\begin{array}{lc|lc}

(\w{eq}_1)                                                         
&\lets{x_1}{V_1}{\lets{x_2}{V_2}{D}}                              &\quad (\w{eq}_2)  &\lets{x}{V}{D} \equiv D  \\
&\equiv \lets{x_2}{V_2}{\lets{x_1}{V_1}{D}}                                && \\
&\mbox{if }x_1\notin\w{FV}(V_2),x_2\notin\w{FV}(V_1)                &&\mbox{if }x\notin \w{FV}(D)

\end{array}
\]
\begin{center}
{\sc Reduction rule}
\end{center}
\[
\begin{array}{c}
(\beta_{V}^{a}) 
\qquad 
E[\lets{x}{\lambda y^+.D}{E'[@(x,z^+)]}]
\arrow 
E[\lets{x}{\lambda y^+.D}{E'[[z^+/y^+]D]}]

\end{array}
\]
\begin{center}
{\sc Typing}
\end{center}
\[
\begin{array}{cc}

(\w{id}^{a}) \quad \infer{x:A \in \Gamma}
{\Gamma \Givesa x:A} 

&(*^{a})\quad \infer{\Gamma,x:\tname{1}\Gives D:A}
{\Gamma \Givesa \lets{x}{*}{D}:A} \\ \\ 

(\lambda^{a})\quad
\infer{\begin{array}{c}
\Gamma,y^+:A^+ \Givesa D':C \\
\Gamma,x:\tname{A^+\arrow C} \Givesa D:B
\end{array}}
{\Gamma \Givesa \lets{x}{\lambda y^+.D'}{D}:B}

&(@^{a})\quad 
\infer{
\begin{array}{c}
\Gamma \Givesa M:\tname{A^+ \arrow B} \\ 
\Gamma \Givesa N^+:A^+
\end{array}}
{\Gamma \Givesa @(M,N^+):B}

\end{array}
\]}

\caption{The $\lambda$-calculus in administrative form: $\lambda^{a}$}
\label{lambda-anf}
\end{table}

\begin{table}
{\footnotesize
\begin{center}
{\sc Translation in AF}
\end{center}
\[
\begin{array}{c}

\adm{1} = \tname{1} \qquad
\adm{A\arrow B} = \tname{\adm{A}\arrow\adm{B}} \\

\adm{x} = x \qquad
\adm{*}=\lets{x}{*}{x} \qquad
\adm{\lambda x.M} =\lets{x}{\lambda x.\adm{M}}{x} \qquad
\adm{MN}          =@(\adm{M},\adm{N})  

\end{array}
\]
\begin{center}
{\sc Readback}
\end{center}
\[
\begin{array}{c}
\rb{\tname{1}} = 1 \qquad
\rb{\tname{A_1\arrow \cdots \arrow A_k \arrow B}} = \rb{A_{1}} \arrow \cdots \arrow \rb{A_{k}} \arrow \rb{B} \\  

\rb{*} = *\qquad
\rb{(\lambda y^+.D)} = \lambda y^+.\rb{D} \\ 

\rb{x}= x \qquad
\rb{(\lets{x}{V}{D})} = [\rb{V}/x]\rb{D} \qquad
\rb{@(M,N_1,\ldots,N_k)} = \rb{M}\rb{N_{1}} \cdots \rb{N_{k}}

\end{array}
\]}
\caption{Translation in administrative form and readback}\label{adm-rb-translation}
\end{table}

\paragraph{Back and forth between $\lambda$ and $\lambda^a$}
In table \ref{adm-rb-translation}, we introduce a translation of
$\lambda$-terms to administrative forms along with a readback
translation.  Clearly, there are $\lambda^a$-terms which are not
structurally equivalent but are mapped to the same $\lambda$-term by
the readback translation.  For instance, taking:
$V \equiv \lambda z.z$, $M\equiv \lets{x}{V}{@(x,x)}$, and 
$N\equiv\lets{x}{V}{\lets{y}{V}{@(x,y)}}$,
we have that $M\not\equiv N$ in
$\lambda^a$ but $\rb{M}\equiv \rb{N}$ in $\lambda$. Thus we can regard
the administrative forms as {\em notations} for $\lambda$ terms which differ 
in the amount of value sharing. Conversely,
given a $\lambda$-term such as $VV$, we can associate with it an
administrative form $\adm{VV} = @(\lets{x}{V}{x},\lets{y}{V}{y}) \equiv
N$.  Thus the translation $(\adm{~})$ in administrative form makes no effort to {\em share} identical
values.  The translation and
the related readback function form a {\em retraction pair} which is
`compatible' with typing and reduction in a sense that is made formal below.  Thus, we can look at the
$\lambda$-calculus as a {\em retract} of the $\lambda^a$ calculus.

In establishing the simulations between $\lambda$ and $\lambda^a$
we {\em cannot} work directly with the translation $(\adm{~})$.
For instance, taking $M=(\lambda x.x(xy))(\lambda z.z)$, 
we have $M\arrow M'$ in $\lambda$ and  $\adm{M} \not\arrow \adm{M'}$. 
Indeed in $M'$ the term $\lambda z.z$ is duplicated while its translation
is shared in $\adm{M'}$. 
We get around this difficulty by working with the readback operation.

\begin{theorem}[read-back translation, functional case]\label{rb-fun}
The following properties hold.
\begin{enumerate}
\item If $D_1\equiv D_2$ in $\lambda^a$ then $\rb{D_1}\equiv \rb{D_2}$ in $\lambda$.

\item If $M$ is a term in $\lambda$ then $\rb{\adm{M}}\equiv M$.

\item 
If $\Gamma \Gives M:A$ in $\lambda$ then $\adm{\Gamma} \Gives^{\w{am}} \adm{M}:\adm{A}$ in $\lambda^a$.

\item 
If $\Gamma \Givesa D:A$ in $\lambda^a$ then $\rb{\Gamma} \Gives \rb{D}:\rb{A}$ in $\lambda$.

\item 
If  $\Gamma \Givesa D_1:A$, 
 $M_{1} \equiv \rb{D_{1}}$ and $D_1 \arrow D_2$  in 
$\lambda^a$ then $M_1 \tarrow M_2$ in $\lambda$ and $M_2\equiv \rb{D_2}$.

\item If  $\Gamma \Gives^{am} D_1:A$,  $M_1\equiv \rb{D_1}$ and $M_1 \arrow M_2$ in $\lambda$  then 
$D_1\arrow D_2$ in $\lambda^{am}$ and  $M_2 \equiv \rb{D_2}$. 

\end{enumerate}
\end{theorem}

The first property states that the readback translation is invariant under
structural congruence and the second that it is the left inverse of the
function that puts a $\lambda$-term in administrative form.
The third and fourth properties state that typing is preserved by the translations.
The fifth property shows that any reduction in $\lambda^a$ corresponds to 
a {\em positive} number of reductions of the readback in $\lambda$ (hence the following corollary).
Finally, the sixth property guarantees that the monadic administrative forms are indeed enough
to simulate the $\lambda$-calculus.

\begin{corollary}\label{adm-sn-corollary}
The $\lambda^a$-calculus terminates.
\end{corollary}

\paragraph{Administrative forms in CPS style}
In table \ref{lambda-anf-cps} we consider a restriction of 
the syntax of the terms where we drop variables (functions are always
applied to their arguments) and the operator $@$ is only 
applied to variables.  
Evaluation contexts are then restricted
accordingly. 
The definition of structural congruence  and reduction are
inherited from $\lambda^a$ and are omitted.
We restrict the syntax of types too by requiring, as it is common in CPS translations, that there is a fixed type of results called $R$.
The resulting language is called $\lambda^{ak}$ and it is a subsystem 
of $\lambda^{a}$ which inherits from $\lambda^{a}$
the reduction and typing rules. In particular, the terms in $\lambda^{ak}$
terminate because those in $\lambda^a$ do.

\begin{table}
{\footnotesize
\begin{center}
{\sc Restricted CPS Syntax}
\end{center}
\[
\begin{array}{lll}

V &::= * \Alt (\lambda \w{id}^{+}.D)                              &\mbox{(values)}\\
D &::= \letv{\w{id}}{V}{M}                                       &\mbox{(declarations)} \\
M &::= @(\w{id},\w{id}^{+})                                      &\mbox{(terms)} \\
E &::= \letv{\w{id}}{V}{[~]}                                     &\mbox{(eval. contexts)}  \\
A &::= \tname{1} \Alt \tname{A^+\arrow R}                         &\mbox{(types)}

\end{array}
\]
\begin{center}
{\sc Specialized typing rules}
\end{center}
\[
\begin{array}{c}


(*^{a})\quad \infer{\Gamma,x:\tname{1}\Gives D:R}
{\Gamma \Givesa \lets{x}{*}{D}:R} 

\qquad
(@^{a})\quad 
\infer{
x:\tname{A^+ \arrow R}, y^+:A^+ \in \Gamma}
{\Gamma \Givesa @(x,y^+):R}
\\
(\lambda^{a})\quad
\infer{
\Gamma,y^+:A^+ \Givesa D':R \quad
\Gamma,x:\tname{A^+\arrow R} \Givesa D:R
}
{\Gamma \Givesa \lets{x}{\lambda y^+.D'}{D}:R} 

\end{array}
\]
}
\caption{Administrative forms in CPS style: $\lambda^{ak}$}
\label{lambda-anf-cps}
\end{table}

\paragraph{CPS translation}
In table \ref{cps-translation}, we describe 
a CPS translation of the administrative language $\lambda^a$ into
$\lambda^{ak}$. The reader familiar with CPS translations, may
appreciate the fact that the translation has been optimized so as 
to have a simple statement and proof of the simulation property. 
In particular, notice that the case for application is
split into two cases. 

\begin{table}
{\footnotesize
\[
\begin{array}{ll}

\cpst{\tname{1}} &= \tname{1} \\

\cpst{\tname{A_1\arrow \cdots \arrow A_n \arrow B}} 
&=\tname{\cpst{A_{1}} \arrow \cdots \arrow \cpst{A_{n}} \arrow K(B) \arrow R}  \\ 
&\mbox{where: }K(B)= \tname{\cpst{B}\arrow R}
\\ \\ 

\psi(*)  &=* \\

\psi(\lambda y^+.D) &=\lambda y^+.\lambda k.(D: k) \\

\letv{x}{V}{M}: k
          &= \letv{x}{\psi(V)}{M:k} \\

@(x^*,@(M,M^+),N^*):k 
&= \lets{k'}{\lambda y.@(x^*,y,N^*): k}{@(M,M^+): k'}  \\ 

@(x,x^+): k &= @(x,x^+,k) \\

x:k       &=@(k,x)

\end{array}\]
}
\caption{An optimized CPS translation on the administrative forms}\label{cps-translation}
\end{table}

\begin{theorem}[CPS translation, functional case]\label{cps-fun}
The following properties hold.

\begin{enumerate}

\item If $\Gamma\Givesa D:A$ then $\cpst{\Gamma},k:K(A) \Givesa (D: k) :R$.

\item If $\Gamma \Givesa D:A$ and  $D\arrow D'$  in $\lambda^a$ then $(D : k) \tarrow (D': k)$ 
in $\lambda^{ak}$.

\end{enumerate}
\end{theorem}

\section{Parallel and Concurrent extensions}\label{conc-sec}
In this section, we explore the more general concurrent case, 
namely the lower part of table
\ref{overview-table}.

\paragraph{A parallel $\lambda$-calculus}
In table \ref{lambda-def-par}, we introduce a parallel version of the $\lambda$-calculus (cf. appendix \ref{spec-app}, table \ref{lambda-def-par-full}).
This amounts to introduce a binary parallel composition operator on terms
along with  a special {\em behavior} type $b$ 
which is attributed to terms running in parallel.
The reduction rule and the typing rules $(\w{id})$, $(*)$, $(\lambda)$, 
$(@)$ are omitted since they are similar to the ones in table
\ref{lambda-def}. Terms running in 
parallel are not supposed to return  a value. The typing guarantees
that  they  cannot occur under an application 
(neither as a function nor as an argument).
Notice that in this language terms running in parallel are not really competing
for the resources, {\em i.e.}, the value declarations. This is because
values are always available and stateless (for this reason we call this
calculus parallel rather than concurrent).
An interesting remark is that the termination 
of the parallel $\lambda$-calculus 
can be derived from the termination of the simply typed $\lambda$-calculus.

\begin{proposition}\label{lambda-par-sn}
The $\lambda_\pr$-calculus terminates.
\end{proposition}

\begin{table}
{\footnotesize
\begin{center}
{\sc Syntax}
\end{center}
\[
\begin{array}{lll}

V &::= * \Alt (\lambda \w{id}.M) \Alt \w{id} &\mbox{(values)}\\
M &::=  V \Alt (MM) \Alt (M \mid M)          &\mbox{(terms)} \\
E &::= [~] \Alt E[[~]M \Alt E[V[~]] \Alt E[[~]\mid M] \Alt E[M\mid [~]]  \quad &\mbox{(eval. contexts)} \\
A &::= 1 \Alt (A\arrow \alpha)               &\mbox{(value types)} \\
\alpha&::= A \Alt b                          &\mbox{(types)} \\
\end{array}
\]
\begin{center}
{\sc New typing rule}
\end{center}
\[
\begin{array}{c}

(\mid)\quad
\infer{\Gamma \Gives M_i:b\quad i=1,2}
{\Gamma \Gives (M_1 \mid M_2):b}

\end{array}
\]}
\caption{Sketch of a parallel $\lambda$-calculus: $\lambda_\pr$}\label{lambda-def-par}
\end{table}

\paragraph{A concurrent $\lambda$-calculus in administrative form}
In table \ref{lambda-anf-conc} we sketch an extension of 
the administrative $\lambda$-calculus 
to accommodate the parallelism already introduced in the $\lambda$-calculus (cf. appendix \ref{spec-app}, table \ref{lambda-anf-conc-full}).
Thus once again we introduce parallel terms and a special behavior type $b$.
In order to have a form of concurrency or competition among the parallel threads we
associate a usage $u$ with each declaration. A usage $u$ varies over
the set $\set{\infty, 1,0}$. The (familiar) idea is that
a declaration with usage $\infty$ is always available,
one with usage $1$ can be used at most once, and one with usage
$0$ cannot be used at all. We take the convention 
that when the usage is omitted the intended usage is $\infty$.
We define an operator $\downarrow$ to {\em decrease} usages as follows:
$\downarrow \infty =\infty$, $\downarrow 1 = 0$, and $\downarrow 0$
is undefined. 
Modulo this enrichment of the declarations with usages, 
the structural congruence is defined as in the functional fragment
of the language (table \ref{lambda-anf}) and it is omitted.
Notice that a reduction is possible only if the
usage of the corresponding definition is not $0$ and in this case the
effect of the reduction is to decrease the usage. 
We omit the typing rules $(\w{id}^a)$, $(*^a)$, and $(@^a)$ which
are similar to the ones in table \ref{lambda-anf}.

\begin{table}
{\footnotesize
\begin{center}
{\sc Syntax}
\end{center}
\[
\begin{array}{lll}

V &::= * \Alt (\lambda \w{id}^{+}.D)             &\mbox{(values)}\\
D &::= \mletv{u}{\w{id}}{V}{M}                      &\mbox{(declarations)} \\
M &::= \w{id} \Alt  @(M,M^{+}) \Alt (M\mid M)        &\mbox{(terms)} \\
E &::= \mletv{u}{\w{id}}{V}{[~]} \Alt E[@(\w{id}^*,[~],M^*)] \Alt E[[~]\mid M] \Alt E[M\mid [~]] \quad
  &\mbox{(eval. contexts)} \\
A &::= \tname{1} \Alt \tname{A^+\arrow \alpha}            &\mbox{(value types)} \\
\alpha &::= A \Alt b                                      &\mbox{(types)} 
\end{array}
\]
\begin{center}
{\sc Reduction rule}
\end{center}
\[
\begin{array}{c}
(\beta_{V}^{a}) 
\qquad 
E[\mlets{u}{x}{\lambda y^+.D}{E'[@(x,z^+)]}]
\arrow 
E[\mlets{\downarrow u}{x}{\lambda y^+.D}{E'[[z^+/y^+]D]}]

\quad 
\end{array}
\]
\begin{center}
{\sc New typing rules}
\end{center}
\[
\begin{array}{c}

(\lambda^{a})\quad
\infer{\begin{array}{c}
u\neq 0\quad \Gamma,y^+:A^+ \Givesa D':\alpha' 
\Gamma,x:\tname{A^+\arrow \alpha'} \Givesa D:\alpha
\end{array}}
{\Gamma \Givesa \mlets{u}{x}{\lambda y^+.D'}{D}:\alpha} \\ \\

(\lambda^{a}_{0})\quad
\infer{
V\neq *\quad \Gamma,x:\tname{A^+\arrow \alpha'} \Givesa D:\alpha
}
{\Gamma \Givesa \mlets{0}{x}{V}{D}:\alpha}

\qquad

(\mid^{a}) \quad
\infer{\Gamma \Givesa M_i:b\quad i=1,2}
{\Gamma \Givesa (M_1\mid M_2):b}

\end{array}
\]}

\caption{Sketch of a concurrent $\lambda$-calculus in administrative form: $\lambda^{a}_{\pr}$}
\label{lambda-anf-conc}
\end{table}

In the typing,  we require that in $\mlets{u}{x}{*}{D}$, $u$ is $\infty$.
Also notice that in $\mlets{0}{x}{V}{D}$ we disregard the typing
of the value $V$.
Given a declaration $D$, we can obtain a declaration $D'$ by replacing all the usages with the
usage $\infty$.  It is clear that all reductions $D$ may perform can
be simulated by $D'$. If the transformation must respect typing then
we can just replace the possibly ill-typed values in $\s{let}_0$ by some well-typed value.
Then, as far as termination is concerned, it is
enough to consider the sub-calculus where all usages are $\infty$;
we denote this calculus with $\lambda^a_{\pr,\infty}$. 
The administrative translation and the related readback translation described
in table \ref{adm-rb-translation} are extended to provide a retraction pair
between the $\lambda_{\pr}$ and the $\lambda^a_{\pr,\infty}$ calculi.
The translations are the identity on the behavior type $b$ and 
distribute over parallel compositions (cf. appendix \ref{spec-app},
table \ref{adm-rb-translation-par-full}):

{\footnotesize
\[
\begin{array}{lllll}
\adm{b} &= b\qquad  &\adm{M \mid M'} &=\adm{M}\mid \adm{M'} \qquad  & \mbox{(translation in AF)}\\
\rb{b}  &= b  &\rb{(M_1 \mid M_2)} &= \rb{M_{1}}\mid \rb{M_{2}} &\mbox{(readback)}
\end{array}
\]}

\begin{theorem}[read-back translation, parallel case]\label{rb-conc}
The following properties hold.

\begin{enumerate}
\item If $D_{1}\equiv D_2$ in $\lambda^{a}_{\pr,\infty}$ then $\rb{D_{1}}\equiv \rb{D_{2}}$ in $\lambda$.

\item If $M$ is a term in $\lambda_{\pr}$ then $\rb{\adm{M}} \equiv M$.

\item 
If $\Gamma \Gives M:\alpha$ in $\lambda_{\pr}$ then $\adm{\Gamma} \Gives^{\w{am}} \adm{M}:\adm{\alpha}$ in 
$\lambda^{am}_{\pr}$.

\item  If $\Gamma \Givesa D:\alpha$ in $\lambda^{a}_{\pr,\infty}$ then $\rb{\Gamma} \Gives \rb{D}:\rb{\alpha}$ 
in $\lambda_{\pr}$.

\item 
If   If $\Gamma \Givesa D_1:\alpha$, 
$M_1 \equiv \rb{D_{1}}$ and $D_{1} \arrow D_{2}$  in $\lambda^{a}_{\pr,\infty}$
then $M_{1}\tarrow M_{2}$ in $\lambda_{\pr}$ and $M_{2}\equiv \rb{D_{2}}$.

\item If  $\Gamma \Gives^{am} D_{1}:\alpha$ in $\lambda^{am}_{\pr,\infty}$,  
$M_{1}\equiv \rb{D_{1}}$ and $M_1 \arrow M_2$ in $\lambda_{\pr}$ then 
$D_1\arrow D_2$ in $\lambda^{am}_{\pr,\infty}$ and  $M_2 \equiv \rb{D_2}$.

\end{enumerate}
\end{theorem}

By theorem \ref{rb-conc}(5), 
a reduction of a term in $\lambda^a_{\pr,\infty}$ corresponds to a {\em positive} number of reductions of the
readback in $\lambda_{\pr}$. Hence, by recalling proposition \ref{lambda-par-sn},
we obtain the following corollary (cf. appendix \ref{expressivity-sec} for concurrent programming examples).

\begin{corollary}\label{adm-sn-corollary-par}
The $\lambda^{a}_{\pr}$-calculus terminates.
\end{corollary}

\paragraph{Concurrent $\lambda$-calculus in administrative, CPS form}
In table \ref{lambda-anf-cps-conc}, we introduce the concurrent
$\lambda$-calculus in administrative and CPS form (cf. appendix \ref{spec-app}, table \ref{lambda-anf-cps-conc-full}).
The typing rules $(*^a)$ and $(@^a)$ are omitted as they are similar
to the ones in table \ref{lambda-anf-cps}.
The CPS translation given in table \ref{cps-translation} is extended to 
a translation from $\lambda^a_{\pr}$ to $\lambda^{ak}_{\pr}$. The behavior
type is mapped to itself, $\cpst{b}=b$, with a continuation type 
$K(b)$ which is conventionally taken to be $\tname{1}$, while
the optimized term translation distributes over parallel composition
$(M\mid M'):k =(M:k) \mid (M':k)$ (cf. appendix \ref{spec-app},
table \ref{cps-translation-conc-full}).
Then typing and reduction are preserved as follows.

\begin{table}
{\footnotesize
\begin{center}
{\sc Restricted CPS Syntax}
\end{center}
\[
\begin{array}{lll}

V &::= * \Alt (\lambda \w{id}^{+}.D)                               &\mbox{(values)}\\
D &::= \mletv{u}{\w{id}}{V}{M}                                        &\mbox{(declarations)} \\
M&::= @(\w{id},\w{id}^{+}) \Alt (M \mid M)                     &\mbox{(terms)} \\
E &::= \mletv{u}{\w{id}}{V}{[~]} \Alt E[[~]\mid M] \Alt E[M\mid [~]]   \quad &\mbox{(eval. contexts)}  \\
A &::= \tname{1} \Alt \tname{A^+\arrow b}                          &\mbox{(types)}

\end{array}
\]
\begin{center}
{\sc New specialized typing rules}
\end{center}
\[
\begin{array}{c}


(\lambda^{a})\quad
\infer{\begin{array}{c}
\Gamma,y^+:A^+ \Givesa D':b \quad
\Gamma,x:\tname{A^+\arrow b} \Givesa D:b
\end{array}}
{\Gamma \Givesa \mlets{u}{x}{\lambda y^+.D'}{D}:b} \\ \\ 

(\lambda^{a}_{0})\quad
\infer{
V\neq *\quad \Gamma,x:\tname{A^+\arrow b} \Givesa D:b
}
{\Gamma \Givesa \mlets{0}{x}{V}{D}:b}

\qquad
(\mid^{a})\quad
\infer{\Gamma \Gives M_i:b\quad i=1,2}
{\Gamma \Gives (M_1 \mid M_2):b}

\end{array}
\]
}
\caption{Sketch of the concurrent administrative forms in CPS style: $\lambda^{ak}_{\pr}$}
\label{lambda-anf-cps-conc}
\end{table}

\begin{theorem}[CPS translation, concurrent case]\label{cps-conc}
The following properties hold.

\begin{enumerate}

\item If $\Gamma\Givesa D:\alpha$ in $\lambda^{a}_{\pr}$ then $\cpst{\Gamma},k:K(\alpha) \Givesa (D: k) :b$ 
in $\lambda^{ak}_{\pr}$.

\item If $\Gamma \Givesa D:\alpha$ and  $D\arrow D'$ in $\lambda^{a}_{\pr}$ then $(D : k) \tarrow (D': k)$ 
in $\lambda^{ak}_{\pr}$.

\end{enumerate}
\end{theorem}

\section{Correspondence with the $\pi$-calculus}\label{pi-sec}
In this section we consider the correspondence between the 
$\lambda^{ak}_{\pr}$-calculus and the $\pi$-calculus.  
To this end, we introduce the syntax of a $\pi$-calculus, $\pi$,
where we write $\{\ldots\}$ to mean that the symbols between curly brackets are optional.
The `functional' version of this calculus (called $\pi_f$ in table \ref{overview-table}) 
is obtained by dropping the (non-replicated) input prefix from the syntax
and it corresponds to the functional administrative CPS forms (table \ref{lambda-anf-cps}).

{\footnotesize
\[
\begin{array}{lll}

D &::= \nu \w{id} \ D \Alt \nu \w{id} \ (\w{id}(\w{id}^+).D \mid D) \Alt \nu \w{id} \ (!\w{id}(\w{id}^+).D \mid D)  \Alt M
                                        &\mbox{(declarations)} \\
M &::= \ol{\w{id}}(\w{id}^{+}) \Alt (M\mid M)                     &\mbox{(terms)} \\
E &::= [~] \Alt E[\nu \w{id}([~])] \Alt E[\nu \w{id}(\{!\}\w{id}(\w{id}^+.D \mid [~]) \Alt 
E[[~]\mid M] \Alt E[M\mid [~]]  \quad &\mbox{(eval. contexts)}\\ 
A &::= \cname{1} \Alt \cname{A^+}                                 &\mbox{(types)}

\end{array}
\]}

The following table provide  a correspondence (a bidirectional translation) between  
$\lambda^{ak}_{\pr}$ and $\pi$. Notice that declarations of the shape $\mlets{\infty}{x}{*}{D}$
and $\mlets{0}{x}{V}{D}$ both correspond to declarations $\nu x \ D$.
The structural congruence, the reduction rules, and the typing rules
of $\pi_r$ are exactly those of $\lambda^{ak}_{\pr}$ modulo this correspondence 
(cf. appendix \ref{spec-app}, table \ref{rigid-pi-full}).

{\footnotesize
\[
\begin{array}{|c|c|c|}  
\hline
&\lambda^{ak}                                                  & \pi \\\hline

\mbox{Types}
&\tname{1}                                                     &\cname{1}  \\
&\tname{A^+ \arrow b}                                          &\cname{A^+}  \\\hline

\mbox{Terms}
&\mlets{\infty}{x}{\lambda y^+.D}{D'}                          &\nu x \ (!x(y^+).D \mid D') \\
&\mlets{1}{x}{\lambda y^+.D}{D'}                               &\nu x \ (x(y^+).D \mid D') \\
&\mlets{0}{x}{V}{D}                                            &\nu x \ D \\
&\mlets{\infty}{x}{*}{D}                                       &\nu x \ D \\
&M \mid M'                                                     &M \mid M' \\
&@(x,y^+)                                                      &\ol{x}y^+ \\\hline

\end{array}
\]}

When applying the translation from $\pi_r$ to $\lambda^{ak}_{\pr}$
it is intended that: (i) $D$ does not contain a (possibly replicated) input prefix
on the declared name $x$ and (ii) the typing determines the appropriate translation
(recall that in $\lambda^{a}_{\pr}$ the typing of $V$ is disregarded).
The reader familiar with the $\pi$-calculus will recognize that
communication is asynchronous (there is no output prefix) and polyadic
(we send vectors of names) and that for every (channel) name $x$ there
is at most one associated definition (a process, possibly replicated,
ready to input on $x$). However these constraints are not enough
to guarantee termination. For instance, consider the 
following looping processes of the (untyped) $\pi$-calculus:
$P_1 \equiv \nu x \ (!x(y).\ol{x}y \mid \ol{x}z)$, 
$P_2 \equiv \nu x,x' \ (!x(y).\ol{x'}y \mid !x'(y).\ol{x}y \mid \ol{x}y)$.
In the presented system, both processes are rejected, 
the first  because the definition of $x$ refers to itself
and the second because the definitions of $x$ and $x'$ are mutually recursive.
The syntactic constraints and the typing rules guarantee 
that the definitions can be linearly ordered so that each definition may only refer
to previously defined names.

\begin{corollary}\label{pi-sn-theorem}
The typed $\pi$-calculus $\pi$ described above terminates (cf. appendix \ref{spec-app}, table \ref{rigid-pi-full}).
\end{corollary}

\section{Conclusion}
We have introduced a simply typed concurrent $\lambda$-calculus in
administrative form, and shown that a fragment of this calculus in
continuation passing style corresponds to a simply typed
$\pi$-calculus. As an application of the correspondence, we have
derived a termination result for the $\pi$-calculus.  
We expect that the current framework can be extended in various
directions including: (i) polymorphic (second order)
types, (ii) refinements towards linear logic/type systems, and (iii) a synchronous/timed variants of the concurrency model.

\newpage

\appendix

\section{Full specifications}\label{spec-app}
This section contains the full specifications of the parallel/concurrent calculi and the
related translations as tables \ref{lambda-def-par-full}, \ref{lambda-anf-conc-full}, 
\ref{adm-rb-translation-par-full}, \ref{lambda-anf-cps-conc-full}, \ref{cps-translation-conc-full}, and
\ref{rigid-pi-full}.

\begin{table}
{\footnotesize
\begin{center}
{\sc Syntax}
\end{center}
\[
\begin{array}{lll}

V &::= * \Alt (\lambda \w{id}.M) \Alt \w{id} &\mbox{(values)}\\
M &::=  V \Alt (MM) \Alt (M \mid M)          &\mbox{(terms)} \\
E &::= [~] \Alt E[[~]M] \Alt E[V[~]] \Alt E[[~]\mid M] \Alt E[M\mid [~]] \quad &\mbox{(eval. contexts)} \\
A &::= 1 \Alt (A\arrow \alpha)               &\mbox{(value types)} \\
\alpha&::= A \Alt b                          &\mbox{(types)} \\
\end{array}
\]
\begin{center}
{\sc Reduction rule}
\end{center}
\[
\begin{array}{c}
(\beta_{V}) \qquad E[(\lambda x.M)V] \arrow E[[V/x]M]
\end{array}
\]
\begin{center}
{\sc Typing}
\end{center}
\[
\begin{array}{cc}

(\w{id})\quad \infer{x:A \in \Gamma}
{\Gamma \Gives x:A} 

&(*)\quad \infer{}
{\Gamma \Gives *:1} \\ \\ 

(\lambda)\quad \infer{\Gamma,x:A \Gives M:\alpha}
{\Gamma \Gives \lambda x.M:A\arrow \alpha} \quad

&(@)\quad \infer{\Gamma \Gives M:A\arrow \alpha\quad \Gamma \Gives N:A}
{\Gamma \Gives MN:\alpha} \\ \\

(\mid)\quad
\infer{\Gamma \Gives M_i:b\quad i=1,2}
{\Gamma \Gives (M_1 \mid M_2):b}

\end{array}
\]}
\caption{A parallel $\lambda$-calculus: $\lambda_\pr$}\label{lambda-def-par-full}
\end{table}

\begin{table}
{\footnotesize
\begin{center}
{\sc Syntax}
\end{center}
\[
\begin{array}{lll}

V &::= * \Alt (\lambda \w{id}^{+}.D)             &\mbox{(values)}\\
D &::= \mletv{u}{\w{id}}{V}{M}                      &\mbox{(declarations)} \\
M &::= \w{id} \Alt  @(M,M^{+}) \Alt (M\mid M)        &\mbox{(terms)} \\
E &::= \mletv{u}{\w{id}}{V}{[~]} \Alt E[@(\w{id}^*,[~],M^*)] \Alt E[[~]\mid M] \Alt E[M\mid [~]] \quad
  &\mbox{(eval. contexts)} \\
A &::= \tname{1} \Alt \tname{A^+\arrow \alpha}            &\mbox{(value types)} \\
\alpha &::= A \Alt b                                      &\mbox{(types)} 
\end{array}
\]
\begin{center}
{\sc Structural Congruence}
\end{center}
\[
\begin{array}{lc|lc}

(\w{eq}_1)                                                         
&\mlets{u_{1}}{x_1}{V_1}{\mlets{u_{2}}{x_2}{V_2}{D}}                              &\quad (\w{eq}_2)  &\mlets{u}{x}{V}{D} \equiv D  \\
&\equiv \mlets{u_{2}}{x_2}{V_2}{\mlets{u_{1}}{x_1}{V_1}{D}}                                && \\
&\mbox{if }x_1\notin\w{FV}(V_2),x_2\notin\w{FV}(V_1)                &&\mbox{if }x\notin \w{FV}(D)

\end{array}
\]
\begin{center}
{\sc Reduction rule}
\end{center}
\[
\begin{array}{c}
(\beta_{V}^{a}) 
\qquad 
E[\mlets{u}{x}{\lambda y^+.D}{E'[@(x,z^+)]}]
\arrow 
E[\mlets{\downarrow u}{x}{\lambda y^+.D}{E'[[z^+/y^+]D]}]

\quad 
\end{array}
\]
\begin{center}
{\sc Typing}
\end{center}
\[
\begin{array}{cc}

(\w{id}^{a}) \quad \infer{x:A \in \Gamma}
{\Gamma \Givesa x:A} 

&(*^{a})\quad \infer{\Gamma,x:\tname{1}\Gives D:\alpha}
{\Gamma \Givesa \mlets{\infty}{x}{*}{D}:\alpha} \\ \\ 

(\lambda^{a})\quad
\infer{\begin{array}{c}
u\neq 0\quad \Gamma,y^+:A^+ \Givesa D':\alpha' \\
\Gamma,x:\tname{A^+\arrow \alpha'} \Givesa D:\alpha
\end{array}}
{\Gamma \Givesa \mlets{u}{x}{\lambda y^+.D'}{D}:\alpha}

&(\lambda^{a}_{0})\quad
\infer{
V\neq *\quad \Gamma,x:\tname{A^+\arrow \alpha'} \Givesa D:\alpha
}
{\Gamma \Givesa \mlets{0}{x}{V}{D}:\alpha}

\\ \\ 

(@^{a})\quad 
\infer{
\begin{array}{c}
\Gamma \Givesa M:\tname{A^+ \arrow \alpha} \\ 
\Gamma \Givesa N^+:A^+
\end{array}}
{\Gamma \Givesa @(M,N^+):\alpha} 

&(\mid^{a}) \quad
\infer{\Gamma \Givesa M_i:b\quad i=1,2}
{\Gamma \Givesa (M_1\mid M_2):b}

\end{array}
\]}

\caption{A concurrent $\lambda$-calculus in administrative form: $\lambda^{a}_{\pr}$}
\label{lambda-anf-conc-full}
\end{table}

\begin{table}
{\footnotesize
\begin{center}
{\sc Translation in AF}
\end{center}
\[
\begin{array}{c}

\adm{1} = \tname{1} \qquad
\adm{b} = b\qquad
\adm{A\arrow \alpha} = \tname{\adm{A}\arrow\adm{\alpha}} \\ \\

\adm{x} = x \qquad
\adm{*}=\mlets{\infty}{x}{*}{x} \qquad
\adm{\lambda x.M} =\mlets{\infty}{x}{\lambda x.\adm{M}}{x} \\
\adm{MN}          =@(\adm{M},\adm{N}) \qquad
\adm{M \mid M'} =\adm{M}\mid \adm{M'}

\end{array}
\]
\begin{center}
{\sc Readback}
\end{center}
\[
\begin{array}{c}
\rb{\tname{1}} = 1 \qquad
\rb{b} = b \qquad
\rb{\tname{A_1\arrow \cdots \arrow A_k \arrow \alpha}} = \rb{A_{1}} \arrow \cdots \arrow \rb{A_{k}} \arrow \rb{\alpha} \\ \\ 

\rb{*} = *\qquad
\rb{(\lambda y^+.D)} = \lambda y^+.\rb{D} 
\qquad
\rb{x}= x  \\  
\rb{(\mlets{\infty}{x}{V}{D})} = [\rb{V}/x]\rb{D} \qquad
\rb{@(M,N_1,\ldots,N_k)} = \rb{M}\rb{N_{1}} \cdots \rb{N_{k}} \\
\rb{(M_1 \mid M_2)} = \rb{M_{1}}\mid \rb{M_{2}}

\end{array}
\]}
\caption{Translation in administrative form and readback: parallel case}\label{adm-rb-translation-par-full}
\end{table}

\begin{table}
{\footnotesize
\begin{center}
{\sc Restricted CPS Syntax}
\end{center}
\[
\begin{array}{lll}

V &::= * \Alt (\lambda \w{id}^{+}.D)                               &\mbox{(values)}\\
D &::= \mletv{u}{\w{id}}{V}{M}                                        &\mbox{(declarations)} \\
M&::= @(\w{id},\w{id}^{+}) \Alt (M \mid M)                     &\mbox{(terms)} \\
E &::= \mletv{u}{\w{id}}{V}{[~]} \Alt E[[~]\mid M] \Alt E[M\mid [~]]   \quad &\mbox{(eval. contexts)}  \\
A &::= \tname{1} \Alt \tname{A^+\arrow b}                          &\mbox{(types)}

\end{array}
\]
\begin{center}
{\sc Specialized typing rules}
\end{center}
\[
\begin{array}{cc}


(*^{a})\quad \infer{\Gamma,x:\tname{1}\Gives D:b}
{\Gamma \Givesa \mlets{\infty}{x}{*}{D}:b}

&(\lambda^{a}_{0})\quad
\infer{
V\neq *\quad \Gamma,x:\tname{A^+\arrow b} \Givesa D:b
}
{\Gamma \Givesa \mlets{0}{x}{V}{D}:b} \\ \\

(\lambda^{a})\quad
\infer{\begin{array}{c}
\Gamma,y^+:A^+ \Givesa D':b \\
\Gamma,x:\tname{A^+\arrow b} \Givesa D:b
\end{array}}
{\Gamma \Givesa \mlets{u}{x}{\lambda y^+.D'}{D}:b} 

&(@^{a})\quad 
\infer{
x:\tname{A^+ \arrow b}, y^+:A^+ \in \Gamma}
{\Gamma \Givesa @(x,y^+):b} \\ \\

(\mid^{a})\quad
\infer{\Gamma \Gives M_i:b\quad i=1,2}
{\Gamma \Gives (M_1 \mid M_2):b}

\end{array}
\]
}
\caption{Concurrent administrative forms in CPS style: $\lambda^{ak}_{\pr}$}
\label{lambda-anf-cps-conc-full}
\end{table}

\begin{table}
{\footnotesize
\[
\begin{array}{ll}

\cpst{\tname{1}} &= \tname{1} \\

\cpst{b}         &=b \\

\cpst{\tname{A_1\arrow \cdots \arrow A_n \arrow \alpha}} 
&=\tname{\cpst{A_{1}} \arrow \cdots \arrow \cpst{A_{n}} \arrow K(\alpha) \arrow b}  \\ 
&\mbox{where: }K(A)=\tname{\cpst{A}\arrow b}, K(b)=\tname{1} \\ \\

\psi(*)  &=* \\

\psi(\lambda y^+.D) &=\lambda y^+.\lambda k.(D: k) \\

\mletv{u}{x}{V}{D}: k
          &= \mletv{u}{x}{\psi(V)}{D:k} \\

(M\mid M'):k &=(M:k) \mid (M':k) \\

@(x^*,@(M,M^+),N^*):k 
&= \mlets{u}{k'}{\lambda y.@(x^*,y,N^*): k}{@(M,M^+): k'}\quad (u\neq 0)  \\ 

@(x,x^+): k &= @(x, x^+,k) \\
x:k       &=@(k,x)

\end{array}\]
}
\caption{An optimized CPS translation for the concurrent case}\label{cps-translation-conc-full}
\end{table}

\begin{table}
{\footnotesize
\begin{center}
{\sc Syntax}
\end{center}
\[
\begin{array}{lll}

D &::= \nu \w{id} \ D \Alt \nu \w{id} \ (\w{id}(\w{id}^+).D \mid D) \Alt \nu \w{id} \ (!\w{id}(\w{id}^+).D \mid D)  \Alt M
                                        &\mbox{(declarations)} \\
M &::= \ol{\w{id}}(\w{id}^{+}) \Alt (M\mid M)                     &\mbox{(terms)} \\
E &::= [~] \Alt E[\nu \w{id}([~])] \Alt E[\nu \w{id}(\{!\}\w{id}(\w{id}^+.D \mid [~]) \Alt 
E[[~]\mid M] \Alt E[M\mid [~]] \quad  &\mbox{(eval. contexts)}\\ 
A &::= \cname{1} \Alt \cname{A^+}                                 &\mbox{(types)}

\end{array}
\]
\begin{center}
{\sc Structural Congruence}
\end{center}
\[
\begin{array}{lc|lc}

(\w{eq}_1)
&\nu x_1 \ (\{\{!\}x_1(y_1^+).D_1 \mid \} \nu x_2 \ (\{\{!\}x_2(y_2^+).D_2 \mid \} D)) 

&(\w{eq}_2)
&\nu x \ (\{\{!\}x(y^+).D' \mid \} D)  \\

&\equiv
\nu x_2 \ (\{\{!\}x_2(y_2^+).D_2 \mid \} \nu x_1 \ (\{\{!\}x_1(y_1^+).D_1 \mid \} D)) 
&&\equiv D \\

&\mbox{if }x_1\notin \w{FV}(\lambda y_{1}^{+}.D_1), x_2\notin \w{FV}(\lambda y_2^+.D_2)
&&
\mbox{if }x\notin \w{FV}(D)
\end{array}
\]
\begin{center}
{\sc Reduction Rules}
\end{center}
\[
\begin{array}{ccc}
E[\nu x \ (!x(y^+).D \mid E'[\ol{x}z^+])] &\arrow
&E[\nu x \ (!x(y^+).D \mid E'[[z^+/y^+]D])] \\

E[\nu x \ (x(y^+).D \mid E'[\ol{x}z^+])] &\arrow
&E[\nu x \ (E'[[z^+/y^+]D])] 

\end{array}
\]
\begin{center}
{\sc Typing Rules}
\end{center}
\[
\begin{array}{lclc}

(\nu^{\pi})&
\infer{\Gamma,x:A \Gives D}{\Gamma \Givespi \nu x\ D}

&(\nu-\w{in}^{\pi})
&
\infer{
\begin{array}{c}
\Gamma,y^+:A^+ \Givespi D' \\
\Gamma,x:\cname{A^+} \Givespi D
\end{array}}
{\Gamma \Givespi \nu x \ (\{!\}x(y^+).D' \mid D)} \\ \\

(\w{out}^{\pi})
&\infer{x:\cname{A^+}, y^+:A^+ \in \Gamma}
{\Gamma \Givespi \ol{x}y^+} 

\quad
&(\mid^{\pi})
&\infer{\Gamma \Givespi M_i\quad i=1,2}
{\Gamma \Givespi (M_1 \mid M_2)}

\end{array}
\]
}
\caption{A concurrent $\pi$-calculus: $\pi$}\label{rigid-pi-full}
\end{table}

\newpage

\section{Expressivity}\label{expressivity-sec}
This section illustrates the expressivity of the $\lambda_{\pr}^{a}$-calculus 
(and therefore of the related $\lambda_{\pr}^{ak}$ and $\pi$-calculi)
as far as the programming of some familiar concepts in concurrent programming 
is concerned. 

\subsection{Output prefix}
An `output prefix' $@(x,y).D$ is simulated by the usual continuation passing
trick:
\[
\begin{array}{lll}
\s{let}_{1}      &k =\lambda w.D   &\s{in} \\
                 &@(x,y,k)
\end{array}
\]

\subsection{Internal choice}
We introduce an `internal choice' operator $\isum$ by defining
$M\isum N$ as follows (all variables being fresh):
\[
\begin{array}{lll}
\s{let}            &x=*                 &\s{in} \\
\s{let}_{1}       &y=\lambda k.@(@(k,y),x)  &\s{in} \\
\s{let}_{1}       &k_1=\lambda w.M  &\s{in} \\
\s{let}_{1}       &k_2=\lambda w.N  &\s{in} \\
&(@(y,k_1) \mid @(y,k_2)) 
\end{array}
\]

\subsection{External choice}
An `external choice' between  $M$ and $N$ based on a boolean value (coded as a projection)
can be defined as follows:
\[
\begin{array}{lll}
\s{let}           &x=*                                            &\s{in} \\
\s{let}_{1}       &y=\lambda z.@(@(z,\lambda w.M,\lambda w.N),x)  &\s{in}  \\
\cdots           
\end{array}
\]

\subsection{Multiple definitions}
One can add to the language the possibility of having multiple definitions of the 
same name.
\[
\begin{array}{lll}
\s{let}          &x=V_1   &\s{or} \\
                 &\cdots    &\s{or} \\
                 &x=V_n   &\s{in} \cdots
\end{array}
\]
where $V_1,\ldots,V_n$ do not depend on $x$.
This does not compromise termination because a multiple definition 
can be simulated by a unique definition that receives its arguments
and then performs an internal choice among the $n$ branches.

\subsection{Joined definitions}
One can also add to the language the possibility of having joined definitions
(in the direction of Fournet and Gonthier join-calculus). 
\[
\begin{array}{lll}
\s{let}          &x_1=V_1   &\s{join} \\
                 &\cdots    &\s{join} \\
                 &x_n=V_n   & \s{in} \cdots
\end{array}
\]
where $V_i$ can only depend on $x_1,\ldots,x_{i-1}$.
The intended semantics is that the definitions of $x_1,\ldots,x_n$ can 
be used only simultaneously. Clearly, a joined definition can be simulated
by a usual one and thus the termination property is not compromised.

\subsection{Lock/Unlock}
One can use the joined definitions to define a lock/unlock mechanism:
\[
\begin{array}{lll}
\s{let} &x= *                     &\s{in} \\
\s{let} &\w{unlock}=\lambda w.\cdots   &\s{join} \\
\s{let} &\w{lock} = \lambda k.@(k,\w{unlock})    &\s{in} \\     
                 & (@(\w{unlock},x) \mid M[\w{lock}])
\end{array}
\]
Here $M[\w{lock}]$ is composed of several threads that may invoke
the $\w{lock}$ definition. When the lock is acquired the thread
receives the name \w{unlock} and invoking it amounts to release the lock
(this is a rudimentary mechanism and no effort is made to 
to enforce a correct usage).

\subsection{CCS channel manager}
Another possible use of the joined definitions is to define a CCS channel manager:
\[
\begin{array}{lll}
\s{let} &x= *                     &\s{in} \\
\s{let} &\w{in} = \lambda k.@(k,x)    &\s{join} \\     
\s{let} &\w{out} = \lambda k.@(k,x)    &\s{in} \\   
                 & M[\w{in},\w{out}]
\end{array}
\]
Here $M$ is composed of parallel threads trying to synchronize on a channel.


\section{Proofs}\label{proofs-sec}
The functional case being a special case of the concurrent one, we
focus directly on the proofs of the latter. There is one
exception: in the concurrent case in administrative, CPS form we fix the type of results to be the
behavior type $b$ rather than an arbitrary (value) type $R$ but this does not
affect the structure of the proofs.

\subsection{Proof of proposition \ref{lambda-par-sn}}
The simple idea idea is to simulate the $\lambda_{\pr}$-calculus
in an ordinary simply typed $\lambda$-calculus equipped with 
a distinguished variable $p$ of type $(b\arrow b)\arrow b$.
More precisely, let $\lambda_p$ be a simply typed $\lambda$-calculus
with two basic types $1$ and $b$, a constant $*$, and 
a distinguished variable $p$. 
It is well-known that such calculus terminates under an arbitrary reduction
strategy. For our purposes, it suffices to consider a reduction strategy
where a call-by-value redex $(\lambda x.M)V$ is reduced in a context
that does not cross a $\lambda$.

Next let us define a translation $\la \_ \ra$ from $\lambda_{\pr}$ to 
to $\lambda_p$ which is the identity on types ($\la \alpha \ra = \alpha$) 
and type contexts and commutes with all the operators of the terms but on parallel composition
where it is defined as follows:
\[
\la M \mid N \ra = (p \la M \ra )\la N \ra~.
\]
The translation is also extended to evaluation contexts where in particular:
\[
\la E[[~]\mid M] \ra = \la E \ra [(p[~])\la M\ra] \quad
\la E[M \mid [~]] \ra = \la E \ra [(p\la M\ra)[~]]
\]
Then it is easy to check the following properties:
\begin{enumerate}

\item If $\Gamma \Gives M:\alpha$ then $\la \Gamma \ra,p:b\arrow (b\arrow b) \Gives \la M \ra : \alpha$.

\item $\la [V/x]M \ra = [\la V \ra/x] \la M \ra$.

\item  $\la E[M] \ra = \la E \ra [\la M \ra]$.

\end{enumerate}

It follows that if $M \arrow N$ in $\lambda_{\pr}$ then $\la M \ra \arrow \la N \ra$ in $\lambda_p$.
Thus since $\lambda_p$ terminates, $\lambda_{\pr}$ must terminate too. \qed

\subsection{Proof of theorem \ref{rb-conc}}

\Proofitem{(1)}
As a preliminary remark notice that if $V$ is a value and $D$ a declaration in $\lambda^a_{\pr}$ then
$\w{FV}(\rb{V})\subseteq \w{FV}(V)$ and 
$\w{FV}(\rb{D})\subseteq \w{FV}(D)$.
Then we proceed by case analysis on the structural congruence by applying the
properties of substitutions.

\Proofitem{(2)} 
By induction on the structure of $M$ term of $\lambda_{\pr}$.

\Proofitem{(3)}
By induction on the typing of $\Gamma\Gives M:\alpha$ in $\lambda_{\pr}$.
For instance, suppose we derive $\Gamma\Gives \lambda x.M:A\arrow \alpha$
from $\Gamma,x:A \Gives M:\alpha$. 
Then by inductive hypothesis, $\adm{\Gamma},x:\adm{A}\Gives^{am} \adm{M}:\adm{\alpha}$.
Also, $\adm{\Gamma},x: \adm{A\arrow \alpha} \Gives^{am} x:\adm{A\arrow \alpha}$,
where by definition $\adm{A\arrow \alpha} = \tname{\adm{A}\arrow \adm{\alpha}}$.
Then we can conclude
$\adm{\Gamma}\Gives^{am} \lets{x}{\lambda x.\adm{M}}{x}:\adm{A\arrow \alpha}$ as required.

\Proofitem{(4)}
First, we prove a substitution lemma for $\lambda_{\pr}$, namely:
if $\Gamma,x:A\Gives M:\alpha$ and $\Gamma \Gives V:A$ then 
$\Gamma \Gives [V/x]M:\alpha$. Then we proceed by
induction on the typing of $\Gamma \Gives D:\alpha$ in $\lambda^{a}_{\pr,\infty}$.

\Proofitem{(5)}
Let $F$ and $E$ be one hole contexts composed of parallel compositions and applications, respectively:
\[
\begin{array}{llll}
F &::= [~] \Alt (F\mid M) \Alt (M\mid F),\qquad
&E&::= [~] \Alt  @(\w{id}^*,E,M^*)~.
\end{array}
\]
If $D_1$ is well-typed and reduces then it has the shape:
\[
\letv{x}{V}{F[E[@(x,z^+)]]}
\]
and $x$ is associated with a value $\lambda y^+.D'$.
Then $D_1 \arrow D_2$ where:
\[
D_2 \equiv \letv{x}{V}{F[E[[z^+/y^+]D']]}~.
\]
We extend the read-back translation to $F$ and $E$ by defining:
\[
\begin{array}{c}
\rb{[~]}=[~] \quad \rb{(F\mid M)} = \rb{F}\mid \rb{M} \qquad \rb{(M\mid F)} = \rb{M} \mid \rb{F} \\
\rb{@(x_1,\ldots,x_n,E,M_1,\ldots,M_m)} = (\cdots (((x_1 \cdots x_n) \rb{E}) \rb{M_{1}}) \cdots \rb{M_{m}}) ~.
\end{array}
\]
Let  $\sigma$ be the (iterated) substitution $[\rb{V}/x]^*$. 
We notice that:
\[
M_1\equiv \rb{D_{1}} = \sigma(\rb{F}[\rb{E}[xz^+]]) = (\sigma \rb{F})[\sigma\rb{E}[\sigma(xz^+)]]~.
\]
Recalling that $\rb{(\lambda y^+.D')} = \lambda y^+.\rb{(D')}$, we have that
$M_1$ performs as many reductions as there are arguments $z^+$ and reduces to 
(assuming suitable renaming of bound variables): 
\[
M_2 \equiv  (\sigma \rb{F})[ (\sigma \rb{E})[ [\sigma z^+/y^+](\sigma \rb{(D')})]~.
\]
On the other hand, we notice that:
\[
\rb{D_{2}} \equiv (\sigma \rb{F})[(\sigma \rb{E})[\sigma([z^+/y^+]\rb{(D')})]]~.
\]
Knowing that the variables $y^+$ do not appear in the domain or codomain of the substitution $\sigma$,
we apply the properties of substitution to check that:
\[
[\sigma z^+/y^+](\sigma \rb{(D')}) \equiv \sigma([z^+/y^+]\rb{(D')})~.
\]

\Proofitem{(6)}
Suppose $D_1$ is typable in the monadic fragment $\lambda^{am}_{\pr}$.
The following table describes how the structure of the readback 
$M_1\equiv \rb{(D_{1})}$ determines $D_1$ up to  structural congruence.
{\footnotesize
\[
\begin{array}{|c|c|}
\hline
M_1    & D_{1}\equiv \\\hline

x   & x \\
*   &\lets{x}{*}{x} \\
\lambda y.M' &\letv{x}{V}{\lets{z}{\lambda y.D'}{z}}  \quad \mbox{ and } \quad M'\equiv \rb{(\letv{x}{V}{D'})}  \\
M_1M_2        &\letv{x}{V}{@(N_1,N_2)} \quad \mbox{ and }\quad M_i\equiv \rb{(\letv{x}{V}{N_i})}, i=1,2 \\
(M_1\mid M_2) &\letv{x}{V}{(N_1 \mid N_2)}\quad \mbox{ and }\quad M_i\equiv \rb{(\letv{x}{V}{N_i})}, i=1,2
 \\\hline

\end{array}
\]}
Suppose $M_1\equiv \rb{D_{1}}$.
Notice that $M_1$ is typable because the readback translation preserves typing (property 4).
Then if $M_1$ reduces, it must have the shape
$M_1=F[E[(\lambda x.M')V]]$, where
$F::= [~] \Alt (F \mid M) \Alt (M \mid F)$ and
$E::= [~] \Alt EM \Alt VE$.
The reduced term is  $M_2=F[E[[V/x]M']]$.
Let $\sigma$ be the (iterated) substitution $[\rb{V}/x]^*$.
By the table above, we derive that $D_1$ must have the shape
$\letv{x}{V}{F'[E'[@(x_1,x_2)]]}$ with: 
\[
\begin{array}{llll}

\sigma \rb{(F')} &= F , 
&\sigma \rb{(E')} &= E, \\
\sigma(x_1) &=\lambda x.M', \quad
&\sigma(x_2) &= V~.

\end{array}
\]
In particular, we see that the variable $x_1$ must occur in the
list of \s{let} declarations and it must be associated with a $\lambda$-abstraction,
say $\lambda x.D'$ where:
\[
\rb{(\letv{x}{V}{D'})}\equiv \sigma \rb{(D')} \equiv M'~.
\]
This means that $D_1$ can also perform one reduction and
reduce to 
\[
D_2\equiv \letv{x}{V}{F'[E'[[x_2/x]D']]}~.
\]
We observe that:
\[
\rb{D_{2}} \equiv (\sigma \rb{(F')})[\sigma \rb{(E')}[\sigma([x_2/x]\rb{(D')})]]
\]
Knowing that the variable $x$  does not appear in the domain or
codomain of the substitution $\sigma$,  we apply the properties
of substitution to check that:
\[
\sigma([x_2/x]\rb{(D')}) \equiv [\sigma(x_2)/x](\sigma \rb{(D')}) \equiv [V/x]M'~.
\]
\qed

\subsection{Proof of theorem \ref{cps-conc}}

\Proofitem{(1)}
By induction on the proof of $\Gamma \Givesa D:\alpha$
and case analysis on the definition of $D:k$.
We spell out the following case: 
\[
\infer{\Gamma \Givesa @(M,M^+):\tname{A^{+}\arrow \alpha}\quad
       \Gamma \Givesa N^+: A^{+}}
{\Gamma \Givesa @(@(M,M^+),N^+):\alpha}
\]
We use the following abbreviations:
\[
\begin{array}{lllllllll}
A'          &\equiv &\tname{A^{+}\arrow \alpha}, \quad
&M'          &\equiv &@(M,M^+), \quad
&\Gamma'     &\equiv  &\cpst{\Gamma},k:K(\alpha)~.
\end{array}
\]
We have to show:
\[
\Gamma' \Givesa \lets{k'}{\lambda y.@(y,N^+):k}{(M':k')}:b
\]
By weakening, we derive  $\Gamma,y:A' \Givesa @(y,N^+):\alpha$.
Then by induction hypothesis we have:
\[
\Gamma', y:\cpst{A'} \Givesa (@(y,N^+): k):b~.
\]
Also, by induction hypothesis  on  $\Gamma \Givesa M':A'$ we derive:
\[
\cpst{\Gamma},k':K(A') \Givesa (M':k'):b~.
\]
Noticing that $K(A') = \tname{\cpst{A'}\arrow b}$ we conclude by
applying the rule for the $\s{let}$. 

\Proofitem{(2)}
As a preliminary remark, we notice that structural congruence is preserved by the CPS translation, 
namely $D\equiv D'$ in $\lambda^a$ entails $(D:k)\equiv (D':k)$.
We introduce some additional notation for evaluation contexts.
We denote with $F$ a one hole context composed of parallel compositions:
\[
F::= [~] \Alt F \mid M \Alt M\mid F~.
\]
We denote with $H$ an elementary applicative context of the shape
$@(\w{id}^*,[~],M^*)$.
If the declaration $D$ in $\lambda^a_{\pr}$ is typable and reduces then
it must have the following shape:
\[
\mletv{u}{x}{V}{F[H_1[\cdots H_m[@(x,z^+)]\cdots]]}
\]
where the name $x$ is associated with a value $\lambda y^+.D'$.
Then the reduced term is:
\[
\mletv{u}{x}{V}{F[H_1[\cdots H_m[[z^+/y^+]D')]\cdots]]}
\]
The CPS translation of the declaration $D$ has the following shape:
\[
\mletv{u}{x}{\psi(V)}{(F[ H_1[\cdots H_m[@(x,z^+)]\cdots]]:k)}
\]
Recalling that the CPS translation distributes over parallel composition,
we define:
\[
[~]:k = [~], \quad  (F\mid M):k = (F:k)\mid (M:k),  \quad (M\mid F):k = (M:k)\mid (F:k)~.
\]
Then we have:
\[
\begin{array}{lll}
F[H_1[\cdots H_m[@(x,z^+)]\cdots]]:k    &= (F:k)[ &\s{let}_{1} \ k_1=\lambda y.(H_1[y]:k) \ \s{in} \\
                                        &         &\s{let}_{1} \ k_2 = \lambda y.(H_2[y]:k_1) \ \s{in}  \\
                                        &         &\cdots \\
                                        &         &\s{let}_{1} \ k_m =  \lambda y.(H_m[y]:k_{m-1}) \ \s{in}  \\
                                        &         &@(x,z^+,k_m) \ ]
\end{array}
\]
Recalling that $\psi(\lambda y^+.D')=\lambda y^+,k.(D':k)$, we observe that
the CPS translation is ready to perform the corresponding reduction.
Then we proceed by case analysis on the shape of $D'$ to show that
the CPS translation reduces to:
\[
\mletv{u}{x}{\psi(V)}{(F[ H_1[\cdots H_m[ [z^+/y^+]D']\cdots]]:k)}
\]

\begin{enumerate}

\item $D'\equiv \mletv{u'}{x'}{V'}{x'}$. 
Then the CPS translation performs a first reduction by replacing 
$@(x,z^+,k_m)$ with 
\[
[z^+/y^+,k_m/k](\mletv{u'}{x'}{\psi(V')}{@(k,x')})~.
\]
If there is no enclosing elementary evaluation context, {\em i.e.} $m=0$, $k_m=k$, 
we are done. Otherwise, the CPS translation performs an additional
reduction along the continuation $k_m$. 
Following this reduction the \s{let} definition of $k_m$ can be removed
applying the second rule of structural congruence.

\item $D'\equiv \mletv{u'}{x'}{V'}{@(M',M'^+)}$.
Then the CPS translation performs a reduction replacing
$@(x,z^+,k_m)$ with 
\[
[z^+/y^+,k_m/k](\mletv{u'}{x'}{\psi(V')}{(@(M',M'^+):k)})~.
\]

\item $D'\equiv \mletv{u'}{x'}{V'}{(M_1\mid M_2)}$.
In this case, the typing requires that there are no enclosing elementary
evaluations contexts, {\em i.e.} $m=0$, $k_m=k$.
The CPS translation performs a reduction replacing
$@(x,z^+,k_m)$ with 
\[
[z^+/y^+,k_m/k](\mletv{u'}{x'}{\psi(V')}{(M_1:k)\mid (M_2:k)})~.
\]

\end{enumerate}

{\footnotesize
\subsection*{Acknowledgment}
The author acknowledges the financial support of the Future and
Emerging Technologies (ET) program within the Seventh Framework
Programme for Research of the European Commission, under FET-Open grant
number: 243881 (project CerCo).}

\end{document}